\documentclass[12pt]{article}
\usepackage[english]{babel}
\usepackage{amssymb,latexsym,amsmath,amscd}

\textwidth=15cm\textheight=23cm\voffset=-12mm\hoffset=-8mm \numberwithin{equation}{section}
\newtheorem{theorem}{Theorem}

\begin{document}

\title
{Classification of BPS equations in higher dimensions}
\author
{E.K. Loginov\footnote{{\it E-mail address:} ek.loginov@mail.ru}\\
\it Department of Physics, Ivanovo State University\\
\it Ermaka St. 39, Ivanovo, 153025, Russia}
\date{}
\maketitle

\begin{abstract}
We systematically classify all possible Bogomol'nyi-Prasad-Sommerfield (BPS) equations in
Euclidean dimension $d\leq8$. We discuss symmetries of BPS equations and their connection with
the self-dual Yang-Mills equations. Also, we present a general method allowing to obtain the
BPS equations in any dimension. In addition, we find all BPS equations in the Minkowski space
of dimension $d\leq6$ and apply the obtained results to the supersymmetric Yang-Mills
theories. In conclusion, we discuss the possibility of using the classification to construct
soliton solutions of the low-energy effective theory of the heterotic string.
\end{abstract}

\section*{Introduction}

Bogomol'nyi-Prasad-Sommerfield (BPS) states are the most important ingredients for recent
developments in nonperturbative aspects of supersymmetric Yang-Mills theory, string theory and
M-theory. In dimension higher that four, BPS configurations can be found as solutions to
first-order equations, known as generalized self-duality or generalized self-dual Yang-Mills
equations. Already more than 20 years ago such equations were proposed~[1,2], and some of
their solutions were found in~[3--13]. In the low energy effective theory, the BPS states
where described by various classical solitonic solutions of various superstring
theories~[14--23]. More recently, various BPS solutions to the noncommutative Yang-Mills
equations in higher dimensions have been investigated in~[24--33].
\par
The main purpose of this paper is to systematically classify possible BPS equations in
Euclidean dimension $d\leq8$. In particular, we consider the super Yang-Mills theories on
Euclidean space, which may be obtained by a dimensional reduction of the $D=10$ $N=1$ super
Yang-Mills theory. In Euclidean dimensions, these theories are realized as the field theoretic
description of $d$ branes. Note that $d$ branes in a background of the Kalb-Ramond field
(NS-NS $B$-field) have been attracting much interest in the development of string theory. The
constant magnetic $B$ field on the $d$ brane, in particular, gives a string theoretical
realization of the non-commutative geometry~[34--36] and the world-volume effective theory on
it is described by the noncommutative Yang-Mills theory. Note also that the $d$-brane bound
states with the $B$ field are very interesting in the context of both brane dynamics and brane
world-volume theory. In the past few years, their systems are discussed from various points of
view in~[37--51].
\par
This paper is organized as follows. In Sec 2, we list the properties of some mathematical
structures relevant to our work. In Sec 3,  we formulate the classified theorem and prove it
in the case of even dimensions. In Secs 4 and 5, we prove the theorem for odd dimensions. In
next section, we present a general method allowing to obtain any systems of BPS equations and
then construct these systems in dimension $d\leq8$. The final section is devoted to
discussions and comments.

\section{Preliminaries}

In this section, we collect the properties of spinors in various dimensions and over $\mathbb
R$ for spaces of various signatures. We also give a brief summary of octonion algebra,
Clifford algebra, and symmetric spaces. We list the features of the mathematical structure as
far as they are of relevance to our work.

\subsection{Spinors}

There are essentially two frameworks for viewing the notion of a spinor. One representation is
theoretic. In this point of view, one knows {\it a priori} that there are some representations
of the Lie algebra of the orthogonal group that cannot be formed by the usual tensor
constructions. These missing representations are then labeled the spin representations, and
their constituents spinors. In this view, a spinor must belong to a representation of the
double cover of the rotation group $SO(d)$, or more generally of the generalized special
orthogonal group $SO(p,q)$ on spaces with metric signature $(p,q)$. These double covers are
Lie groups, called the spin groups $Spin(p,q)$. All the properties of spinors, and their
applications and derived objects, are manifested first in the spin group. The other point of
view is geometrical. One can explicitly construct the spinors, and then examine how they
behave under the action of the relevant Lie groups. This latter approach has the advantage of
being able to say precisely what a spinor is, without invoking some nonconstructive theorem
from representation theory. Representation theory must eventually supplement the geometrical
machinery once the latter becomes too unwieldy. Therefore, we well use the representation
theoretic frameworks for viewing the notion of a spinor.
\par
Let $\mathbb R^{p,q}$ be a finite real space with the nondegenerate metric $\eta$ of signature
$(p,q)$. We choose the orthogonal basis
$\Gamma_1,\dots,\Gamma_{p},\Gamma_{p+1},\dots,\Gamma_{p+q}$ in $\mathbb R^{p,q}$, so as the
quadratic form $\eta$ has the standard diagonal form
\begin{equation}
\eta=\text{diag}(1,\dots,1,-1,\dots,-1).
\end{equation}
Clifford algebra $Cl_{p,q}(\mathbb R)$ is a real associative algebra generated by elements of
$\mathbb R^{p,q}$ and defined by the relations
\begin{equation}\label{12-08}
\Gamma_{a}\Gamma_{b}+\Gamma_{b}\Gamma_{a}=2\eta_{ab}.
\end{equation}
It follows from (\ref{12-08}) that the matrices $\Gamma_{a}$ are unitary if we impose the
conditions
\begin{equation}\label{12-22}
\Gamma_{a}^{\dag}=\Gamma^{a}.
\end{equation}
The algebra $Cl_{p,q}(\mathbb R)$ has dimension $2^{p+q}$, and its element is a linear
combination of the monomials
\begin{equation}\label{12-03}
\Gamma_{a_{1}a_{2}\dots a_{k}}=\Gamma_{a_{1}}\Gamma_{a_{2}}\dots\Gamma_{a_{k}},
\end{equation}
where $1\leqslant a_{1}<a_{2}<\dots<a_{k}\leqslant p+q$. It is obvious that the set of all
monomials (\ref{12-03}) with the identity of $Cl_{p,q}(\mathbb R)$ form its basis. This basis
is called canonical.
\par
The subalgebra of $Cl_{p,q}(\mathbb R)$ generated by all monomials $\Gamma_{ab}$ is called
even and denoted by the symbol $Cl^0_{p,q}(\mathbb R)$. Since
\begin{equation}\label{12-16}
[\Gamma_{ab},\Gamma_{cd}]=\eta_{ad}\Gamma_{bc}+\eta_{bc}\Gamma_{ad}
-\eta_{ac}\Gamma_{bd}-\eta_{bd}\Gamma_{ac},
\end{equation}
its commutator algebra contains the Lie algebra $so(p,q)$. The follows isomorphisms are true:
\begin{align}
Cl^0_{p,q}(\mathbb R)&\simeq Cl_{p,q-1}(\mathbb R),\qquad q>0,\label{12-04}\\
Cl^0_{p,q}(\mathbb R)&\simeq Cl_{q,p-1}(\mathbb R),\qquad p>0.\label{12-05}
\end{align}
Complexifying the vector space $Cl_{p,q}(\mathbb R)$, we get the complex Clifford algebra
$Cl_{d}(\mathbb C)$, where $d=p+q$. This algebra is isomorphic to the algebra $\mathbb
C(2^{n})$ of all complex $2^{n}\times 2^{n}$ matrices, if $d=2n$, or the direct sum of such
algebras, if $d=2n+1$, i.e.
\begin{align}
Cl_{2n}(\mathbb C)&\simeq\mathbb C(2^{n}),\\
Cl_{2n+1}(\mathbb C)&\simeq\mathbb C(2^{n})\oplus\mathbb C(2^{n}).
\end{align}
It therefore has a unique irreducible representation of dimension $2k$. Any such irreducible
representation is, by definition, a space of spinors called a spin representation.
\par
The Pin group $Pin(p,q)$ is the subgroup of the multiplicative group of elements of norm 1 in
$Cl_{p,q}(\mathbb R)$, and similarly the Spin group $Spin(p,q)$ is the subgroup of even
elements in $Pin(p,q)$. It is obvious that any representation of $Cl_{p,q}(\mathbb C)$ induces
a complex representation of $Spin(p,q)$. One is called the Dirac representation. In odd
dimensions, this representation is irreducible. In even dimensions, it is reducible when taken
as a representation of $Spin(p,q)$ and may be decomposed into two: the left-handed and
right-handed Weyl spinor representations. In addition, sometimes the noncomplexified version
of $Cl_{p,q}(\mathbb R)$ has a smaller real representation, the Majorana spinor
representation. If this happens in an even dimension, the Majorana spinor representation will
sometimes decompose into two Majorana-Weyl spinor representations. Of all these, only the
Dirac representation exists in all dimensions. Dirac and Weyl spinors are complex
representations, while Majorana spinors are real representations.
\par
The irreducible representations of $Spin(p,q)$ for $p+q<8$ can be obtained from Table 1, if we
make use of the isomorphisms (\ref{12-04}) and (\ref{12-05}).
\bigskip\par
{\small\noindent Table~1. Representations of the Clifford algebra $Cl_{p,q}(\mathbb R)$
$$
\arraycolsep=0.6mm
\begin{array}{p{6mm}lllllllllllllll}
\hline
&-7&-6&-5&-4&-3&-2&-1&0&1&2&3&4&5&6&7\\
0&&&&&&&&\mathbb R&&&&&&&\\
1&&&&&&&\mathbb C&&\mathbb R^2&&&&&&\\
2&&&&&&\mathbb H&&\mathbb R(2)&&\mathbb R(2)&&&&&\\
3&&&&&\mathbb H^2&&\mathbb C(2)&&\mathbb R^2(2)&&\mathbb C(2)&&&&\\
4&&&&\mathbb H(2)&&\mathbb H(2)&&\mathbb R(4)&&\mathbb R(4)&&\mathbb H(2)&&&\\
5&&&\mathbb C(4)&&\mathbb H^2(2)&&\mathbb C(4)&&\mathbb R^2(4)&&\mathbb
C(4)&&\mathbb H^2(2)&&\\
6&&\mathbb R(8)&&\mathbb H(4)&&\mathbb H(4)&&\mathbb R(8)&&\mathbb R(8)&&\mathbb
H(4)&&\mathbb H(4)&\\
7&\mathbb R^2(8)&&\mathbb C(8)&&\mathbb H^2(4)&&\mathbb C(8)&&\mathbb R^2(8)&&\mathbb
C(8)&&\mathbb H^2(4)&&\mathbb C(8)\\ \hline
\end{array}
$$
Here $p+q$ runs vertically, $p-q$ runs horizontally, and $\mathbb A^2\equiv\mathbb
A\oplus\mathbb A$.}
\bigskip\par\noindent
Table 1 continues with a periodicity of eight, that is, $Cl_{p+8,q}\simeq Cl_{p,q+8}\simeq
Cl_{p,q}(16)$, which is the $16\times16$ matrix algebra with entries in the Clifford algebra
$Cl_{p,q}(\mathbb R)$. Therefore, in fact, we have spinor representations of $Spin(p,q)$ any
$p$ and $q$. For example, the Dirac representation of $Spin(2n+1)$ is real, if $n\equiv
0,3\mod 4$, and pseudoreal, if $n\equiv 1,2\mod 4$.  The Weyl representations of $Spin(2n)$
are complex conjugates of one another as $n\equiv 1\mod 2$, real as $n\equiv 0\mod 4$, and
pseudoreal as $n\equiv 2\mod 4$. These two representations are dual of one another, if $n$ is
odd, and self-dual, if $n$ is even.

\subsection{Octonions}

We recall that the algebra of octonions $\mathbb O$ is a real linear algebra with the
canonical basis $1,e_{1},\dots,e_{7}$ such that
\begin{equation}\label{12-09}
e_{i}e_{j}=-\delta_{ij}+c_{ijk}e_{k},
\end{equation}
where the structure constants $c_{ijk}$ are completely antisymmetric and nonzero and equal to
unity for the seven combinations (or cycles)
$$
(ijk)=(123),(145),(167),(246),(275),(374),(365).
$$
The algebra of octonions is not associative but alternative, i.e. the associator
\begin{equation}\label{12-06}
(x,y,z)=(xy)z-x(yz)
\end{equation}
is totally antisymmetric in $x,y,z$. Consequently, any two elements of $\mathbb O$ generate an
associative subalgebra. The algebra of octonions satisfies the identity
\begin{equation}\label{12-21}
((zx)y)x=z(xyx),
\end{equation}
which is called the right Moufang identity. The algebra $\mathbb O$ permits the involution
(anti-automorphism of period two) $x\to\bar x$ such that the elements
\begin{equation}
t(x)=x+\bar x,\qquad n(x)=\bar xx
\end{equation}
are in $\mathbb R$. In the canonical basis, this involution is defined by $\bar e_{i}=-e_{i}$.
It follows that the bilinear form
\begin{equation}\label{12-07}
(x,y)=\frac12(\bar xy+\bar yx)
\end{equation}
is positive definite and defines an inner product on $\mathbb O$. It is easy to prove that the
quadratic form $n(x)$ permits the composition
\begin{equation}\label{12-23}
n(xy)=n(x)n(y).
\end{equation}
Since the quadratic form $n(x)$ is positive definite, it follows that $\mathbb O$ is a
division algebra. Linearization of (\ref{12-23}) to $x$ and $y$ gives
\begin{align}
n(x)(y,z)&=(xy,xz)=(yx,zx),\label{12-13}\\
2(x,y)(z,t)&=(xz,yt)+(xt,yz).\label{12-14}
\end{align}
Finally, notice that the algebra of octonions is unique, to within isomorphism, alternative
nonassociative simple real division algebra.
\par
Now let $\Gamma_1,\dots,\Gamma_7$ be generators of the Clifford algebra $Cl_{0,7}(\mathbb R)$
satisfying the relations (\ref{12-08}). Further, let $x\in\mathbb O$. Denote by $R_{x}$ the
operator of right multiplication in $\mathbb O$
\begin{equation}\label{12-17}
yR_{x}=yx,\qquad y\in\mathbb O.
\end{equation}
Using the multiplication law (\ref{12-09}) and antisymmetry of the associator (\ref{12-06}),
we prove the equalities
\begin{equation}\label{12-10}
R_{e_{i}}R_{e_{j}}+R_{e_{j}}R_{e_{i}}=-2\delta_{ij}E,
\end{equation}
where $E$ is the identity $8\times 8$ matrix. Comparing (\ref{12-10}) with (\ref{12-08}), we
see that the correspondence $\Gamma_{i}\to R_{e_{i}}$ can be extended to the homomorphism
\begin{equation}\label{12-11}
Cl_{0,7}(\mathbb R)\to\text{End}\,\mathbb O.
\end{equation}
Using Table 1, we prove that the mapping (\ref{12-11}) is surjective and $\text{End}\,\mathbb
O\simeq\mathbb R(8)$. Since
\begin{equation}\label{12-12}
Cl_{0,7}(\mathbb R)\simeq Cl^0_{8,0}(\mathbb R),
\end{equation}
it follows that the homomorphism (\ref{12-11}) induces the homomorphism $Spin(8)\to SO(8)$. We
define the sets
\begin{align}
\mathbb S^7&=\{a\in\mathbb O\mid n(a)=1\},\\
\mathbb S^6&=\{\boldsymbol a\in\mathbb O\mid n(\boldsymbol a)=1\},
\end{align}
where $\boldsymbol a$ is a vector part of the octonion $a=a_0+\boldsymbol a$. It follows from
(\ref{12-11}), (\ref{12-12}), and (\ref{12-13}) that the sets
\begin{align}
X&=\{R_{a}\mid a\in\mathbb S^7\},\label{12-18}\\
Y&=\{R_{\boldsymbol a}R_{\boldsymbol b}\mid \boldsymbol a,\boldsymbol b\in\mathbb
S^6\}\label{12-19}
\end{align}
generate the groups $SO(8)$ and $Spin(7)$, respectivelly. Note also that the product
\begin{equation}\label{12-15}
R_{e_1}R_{e_2}\dots R_{e_7}=E.
\end{equation}
The equality (\ref{12-15}) follows from simplicity of $\mathbb R(8)$ and the fact that the
element $\Gamma_1\Gamma_2\dots\Gamma_7$ lies in the center of $Cl_{0,7}(\mathbb R)$. It
follows from (\ref{12-15}) that restriction of the homomorphism (\ref{12-11}) on $Spin(7)$ is
injection.

\subsection{Symmetric spaces}

We list the properties of symmetric spaces relevant to our work. Let $G$ be a connected Lie
group, $\sigma$ an involutive automorphism of $G$, and $G_{\sigma}$ a set of all fixed point
of $G$ under $\sigma$. Further, let $H$ be a closed subgroup in $G_{\sigma}$ containing the
identity component of $G_{\sigma}$. The quotient space $G/H$ is called a symmetric homogeneous
space. If the subgroup $H$ is compact, then the space $G/H$ admits an $G$-invariant Riemannian
metric. The symmetric space $G/H$ equipped with such metric is called a globally symmetric
Riemannian space.
\par
Automorphism $\sigma$ induces an involutive automorphism of the Lie algebra $A$ of the group
$G$. With respect to this automorphism the algebra $A$ can be decomposable into the direct sum
\begin{equation}\label{12-20}
A=A^{+}\oplus A^{-}
\end{equation}
of proper subspaces corresponding to the eigenvalues $\pm1$. We have obviously
\begin{equation}
[A^{+},A^{+}]\subseteq A^{+},\qquad [A^{+},A^{-}]\subseteq A^{-},\qquad [A^{-},A^{-}]\subseteq
A^{+}.
\end{equation}
The space $A^{+}$ coincides with the Lie algebra of the group $H$, and the space $A^{-}$ is
closed under the composition $[x,y,z]=[[x,y],z]$. The vector space $A^{-}$ equipped with this
trilinear composition is called a triple Lie system.
\par
A globally symmetric Riemannian space $G/H$ is said to be irreducible if the algebra $A$ is
semisimple, the subalgebra $A^{+}$ is a maximal proper subalgebra in $A$, and $A^{+}$ contains
no nonzero ideals of $A$. In particular, irreducible global symmetric Riemannian spaces are
the spaces
\begin{align}
M_{pq}=SO(p+q)&/SO(p)\times SO(q),\\
N_{pq}=SU(p+q)&/S(U(p)\times U(q)).
\end{align}
Note that $M_{pq}$ and $N_{pq}$ are compact simple connected spaces of dimension $pq$ and
$2pq$, respectively.

\section{The main theorem}

BPS states refer to field configurations which are invariant under some supersymmetries. In
super Yang-Mills theories on the Euclidean space $\mathbb R^{d}$, a bosonic configuration is
BPS if there exist a nonzero constant spinor $\varepsilon$ in an unitary space $V$ of
dimension $2^{[d/2]}$, where $[d/2]$ is an integral part of $d/2$, such that the infinitesimal
supersymmetric transformation of the fermion field vanishes
\begin{equation}\label{13-01}
\delta \chi=F_{ab}\Gamma_{ab}\varepsilon=0.
\end{equation}
Such zero eigenspinors of the matrix $F_{ab}\Gamma_{ab}$ form the subspace $W\subseteq V$. The
BPS field strength should satisfy certain conditions in order to have a given number of
unbroken supersymmetries. These conditions can be written as a system of linear equations (BPS
equations) connecting components of $F_{ab}$. We say that two systems of BPS equations are
equivalent if either they are incompatible or they have the same solutions up to a
nondegenerate transformation of $\mathbb R^{d}$. Otherwise, they are called nonequivalent.
Since we consider a global supersymmetry, the conditions imposed on $F_{ab}$ do not depend on
a choice of basis in $\mathbb R^{d}$. Hence, we must find nonequivalent systems of BPS
equations.
\par
In order that to find such systems, we define the projection operator $\Omega$ mapping $V$
onto $W$, as has been done previously in~\cite{bak02}. With an suitable orthonormal basis for
$V$, this operator appears as $2^{[d/2]}\times 2^{[d/2]}$ matrix
\begin{equation}\label{13-07}
\tilde\Omega=\begin{pmatrix} E_{r}&0\\0&0\end{pmatrix},
\end{equation}
where $E_{r}$ is the identity $r\times r$ matrix, and $r=\dim W$. Obviously, the projection
operator is diagonalizable in an orthonormal basis, and it has a real spectrum (its
eigenvalues are 0 or 1). Therefore, it is Hermitian. Thus,
\begin{align}
\Omega^2&=\Omega,\label{11-25}\\
\Omega^{\dag}&=\Omega\label{11-26}.
\end{align}
Now we can rewrite the Eq. (\ref{13-01}) in the following equivalent form
\begin{equation}\label{13-02}
F_{ab}\Gamma_{ab}\Omega=0.
\end{equation}
In order to get system of BPS equations from (\ref{13-02}), we must represent the projector
$\Omega$ as a linear combination of the identity matrix and the monomials (\ref{12-03}), and
further use the identities (\ref{12-08}). Note also that the constant $\nu$, defined by
\begin{equation}
\text{tr}\,\Omega=\nu\times 2^{[d/2]},
\end{equation}
gives the fraction of the unbroken supersymmetry, so $0\leq\nu\leq1$. The $\nu=0$ or 1 cases
are trivial, either meaning the non-BPS state or the vacuum, $F_{ab}=0$. The following theorem
contains the main result of the paper:
\begin{theorem}
Suppose the constant spinor $\varepsilon$ satisfying (\ref{13-01}) is Weyl as even $d$,
Majorana as $d=7$, and Majorana-Weyl as $d=8$. Then there exists to within equivalence a
unique system of BPS equations for every pair of values $d\leq 8$ and $\nu=\nu(d)$.
\end{theorem}
{\bf Proof.} Let $\{e_{a}\}$ and $\{e'_{a}\}$ be two orthonormal bases in $\mathbb R^{d}$.
Then there exists an orthogonal transformation of $\mathbb R^{d}$ such that
$e'_{a}=A^{b}_{a}e_{b}$. In this case, the components of $F_{ab}$ are transformed by the rule
\begin{equation}\label{13-03}
F^{k}_{ab}\to F^{k}_{cd}A^{c}_{a}A^{d}_{b}.
\end{equation}
Denote by the symbol $F^{k}$ a real skew-symmetric $d\times d$ matrix with the elements
$F^{k}_{ab}$. Then the transformation (\ref{13-03}) can be rewritten in the matrix form
\begin{equation}\label{13-04}
F^{k}\to AF^{k}A^{-1},
\end{equation}
where $A$ is an orthogonal matrix with the elements $A_{ab}=A^{b}_{a}$ such that $\det A=1$.
Obviously, the matrices $A$ and $F^{k}$ are elements of the group $SO(d)$ and the algebra
$so(d)$, respectively. Since $F^{k}$ is arbitrary real skew-symmetric matrix, it follows that
the transformation (\ref{13-04}) defines an inner automorphism of $so(d)$.
\par
On the other hand, the antisymmetry matrices $\Gamma_{ab}$ satisfy the commutation relations
(\ref{12-16}). Therefore they generate a Lie algebra $\widetilde{so}(d)$ that is isomorphic to
$so(d)$. Denote by $\tilde F^{k}$ an image of $F^{k}$ with respect to the isomorphism
$so(d)\to\widetilde{so}(d)$. Then we have the following diagram,
\begin{equation}\label{13-05}
\begin{CD}
F^{k} @>>> AF^{k}A^{-1} \\
@VVV @VVV \\
\tilde F^{k} @>>> B\tilde F^{k}B^{-1}
\end{CD},
\end{equation}
\medskip
\par\noindent
where the matrix $B\in Spin(d)$. It is obvious that this diagram is commutative. In
particular, any inner automorphism of $\widetilde{so}(d)$ defined by the mapping
\begin{equation}
\tilde F^{k}\to B\tilde F^{k}B^{-1}
\end{equation}
induces the transformation (\ref{13-03}).
\par
Further, the matrices $\Gamma_{ab}$ make up a basis of $\widetilde{so}(d)$. Therefore, any of
its element $\tilde F^{k}$ can be represented in the form
\begin{equation}
\tilde F^{k}=\tilde F^{k}_{ab}\Gamma_{ab}.
\end{equation}
Denote by $\tilde F_{ab}$ an antisymmetry tensor with the components $\tilde F^{k}_{ab}$ and
consider the equation
\begin{equation}\label{13-06}
\tilde F_{ab}\Gamma_{ab}(B^{-1}\Omega B)=0.
\end{equation}
It follows from commutativity of the diagram that the Eqs. (\ref{13-02}) and (\ref{13-06}) are
equivalent. Thus, if we prove that by the transformation
\begin{equation}\label{13-08}
\Omega\to\tilde\Omega=B^{-1}\Omega B,
\end{equation}
where $B\in Spin(d)$, the matrix $\Omega$ can be reduced to the form (\ref{13-07}), then we
prove the theorem.
\par
We consider even dimension $d=2n$. Without loss of generality, we can suppose that
$\varepsilon$ is a right-handed (chiral) spinor, i.e.
\begin{equation}\label{13-10}
\Gamma_{*}\varepsilon=\varepsilon,\qquad\Gamma_{*}=(-i)^{n}\Gamma_{1}\dots \Gamma_{2n}.
\end{equation}
We will seek representations of gamma matrices such that
\begin{equation}\label{13-11}
\Gamma_{a}=\begin{pmatrix}
0&\Lambda_{a}\\\Lambda_{a}^{\dag}&0\endpmatrix,\qquad\Gamma_{*}=\pmatrix
E&0\\0&-E\end{pmatrix},
\end{equation}
where $E$ is the identity matrix. It is obvious that in this representation, the Hermitian
projection operator $\Omega$ and the element $B$ of $Spin(n)$ take the form
\begin{equation}\label{13-12}
\Omega=\begin{pmatrix}\Omega_{+}&0\\0&0\endpmatrix,\qquad B=\pmatrix
B_{+}&0\\0&B_{-}\end{pmatrix},
\end{equation}
where $\Omega$ and $B_{\pm}$ are $n\times n$ matrices. Now we consider the concrete values of
$n$.
\par
(1) In two Euclidean dimensions, the Weyl spinor is one-component and complex representation.
Therefore, we choose the representation in terms of Pauli matrices
\begin{equation}
\begin{aligned}
\Gamma_1&=\sigma_1,\\
\Gamma_2&=\sigma_2.
\end{aligned}
\end{equation}
It follows from (\ref{11-25}) that $\Omega_{+}=0$ or $1$.
\par
(2) In four Euclidean dimensions, there are two inequivalent pseudoreal two-component Weyl
spinor, and each of them transform under $SU(2)$. We choose the gamma matrices in the form
\begin{equation}
\begin{aligned}
\Gamma_{k}&=\sigma_{1}\otimes\sigma_{k},\\
\Gamma_{4}&=\sigma_{2}\otimes\sigma_{0},
\end{aligned}
\end{equation}
where $\sigma_{0}$ is the identity $2\times2$ matrix. In this representation, the generators
of $Spin(4)$ have the block diagonal form
\begin{equation}
\begin{aligned}
\Gamma_{ij}&=i\varepsilon_{ijk}(\sigma_{0}\otimes\sigma_{k}),\\
\Gamma_{k4}&=i(\sigma_{3}\otimes\sigma_{k}).
\end{aligned}
\end{equation}
The matrices $\sigma_{k}$ form a basis of $su(2)$. Therefore, $B_{+}$ is arbitrary unitary
$2\times2$ matrix. Since the matrix $\Omega$ is Hermitian, if follows that it can be reduced
to the form (\ref{13-07}) by the transformation (\ref{13-08}).
\par
(3) In six Euclidean dimensions, the isomorphism $Spin(6)\simeq SU(4)$ guarantees that there
are two four-dimensional complex Weil representations that are complex conjugates of one
another. We choose the gamma matrices in the form
\begin{equation}
\begin{aligned}
\Gamma_{k}&=\sigma_{1}\otimes\sigma_{k}\otimes\sigma_{0},\\
\Gamma_{k+3}&=\sigma_{2}\otimes\sigma_{0}\otimes\sigma_{k},
\end{aligned}
\end{equation}
where $k=1,2,3$. In this representation, the generators of $Spin(6)$ have the following form
\begin{equation}
\begin{aligned}
\Gamma_{ij}&=i\varepsilon_{ijk}(\sigma_{0}\otimes\sigma_{k}\otimes\sigma_{0}),\\
\Gamma_{i(j+3)}&=i(\sigma_{3}\otimes\sigma_{i}\otimes\sigma_{j}),\\
\Gamma_{(i+3)(j+3)}&=i\varepsilon_{ijk}(\sigma_{0}\otimes\sigma_{0}\otimes\sigma_{k}).
\end{aligned}
\end{equation}
Noting that the matrices $\sigma_{k}\otimes\sigma_{0}$, $\sigma_{i}\otimes\sigma_{j}$, and
$\sigma_{0}\otimes\sigma_{k}$ form a basis of $su(4)$, we prove that $B_{+}$ is arbitrary
unitary $4\times4$ matrix. Hence the Hermitian matrix $\Omega$ can be reduced to the form
(\ref{13-07}) by (\ref{13-08}).
\par
(4) In eight Euclidean dimensions, the Weyl-Majorana representation is eight-dimensional and
real. We choose the $\Gamma$ matrices in the form
\begin{equation}\label{13-09}
\Gamma_{8}=\begin{pmatrix} 0&E\\E&0\end{pmatrix},\qquad
\Gamma_{k}=\begin{pmatrix}0&R_{k}\\-R_{k}&0\end{pmatrix},
\end{equation}
where the real $8\times8$ matrices $R_{k}$ ($k=1,\dots,7$) are antisymmetric and satisfy
\begin{equation}
R_{i}R_{j}+R_{j}R_{i}=-2\delta_{ij}E.
\end{equation}
Obviously, we can choose this matrices in the form of operators (\ref{12-17}) of right
multiplication on the basic elements $e_{i}$ of $\mathbb O$, i.e. we suppose
$R_{i}=R_{e_{i}}$. Since by (\ref{12-15}) the product
\begin{equation}
R_1R_2\dots R_7=E,
\end{equation}
the matrix $\Gamma_{*}$ has the form (\ref{13-11}). It follows from (\ref{13-09}) that the
generators of $Spin(8)$ are
\begin{equation}
\Gamma_{i8}=\begin{pmatrix} R_{i}&0\\0&-R_{i}\end{pmatrix},\qquad \Gamma_{ij}=\begin{pmatrix}
[R_{j},R_{i}]&0\\0&[R_{j},R_{i}]\end{pmatrix}.
\end{equation}
The elements $R_{i}$ and $[R_{j},R_{i}]$ make up an basis of $so(8)$. Therefore, $B_{+}$ is an
arbitrary orthogonal $8\times8$ matrix. Since $\Omega$ is a real symmetric matrix, it can be
reduced to the form (\ref{13-07}) by the transformation (\ref{13-08}).

\section{Seven dimensions}

In seven Euclidean dimensions, the single spinor representation is eight dimensional and real.
Therefore, the projection operator $\Omega$ is represented as $8\times8$ real symmetric
matrix. We must prove that
\begin{equation}\label{14-01}
B^{-1}\Omega B=\tilde\Omega=\text{diag}\{1,\dots,1,0,\dots,0\}
\end{equation}
for some $B\in Spin(7)$. In the first place, we note that there exists an element $U\in SO(8)$
such that
\begin{equation}\label{14-02}
\tilde\Omega=U\Omega U^{-1}
\end{equation}
Then it follows from (\ref{14-01}) and (\ref{14-02}) that
\begin{equation}\label{14-03}
\tilde\Omega\tilde B=\tilde B\tilde\Omega,
\end{equation}
where the matrix $\tilde B=UB$. Further, the general solution of the Eq. (\ref{14-03}) has the
form
\begin{equation}
\tilde B=\begin{pmatrix}\tilde B_1&0\\0&\tilde B_2\end{pmatrix},
\end{equation}
where $\tilde B_1$ and $\tilde B_2$ are orthogonal matrices such that $\det\tilde B_{i}=1$.
Therefore,
\begin{equation}
\tilde B\in H_{k}\simeq SO(k)\times SO(8-k),\qquad 1\leqslant k\leqslant4.
\end{equation}
On the other hand, $U=\tilde BB^{-1}$. Hence the equality (\ref{14-01}) is true if the group
\begin{equation}\label{14-04}
SO(8)=H_{k}Spin(7),
\end{equation}
i.e. if any element $g\in SO(8)$ can be represented as the product $g=hf$, where $h\in H_{k}$
and $f\in Spin(7)$. We will prove the equality (\ref{14-04}).

\subsection{The case $k=1$.}

As stated above, the groups $SO(8)$ and $Spin(7)$ are generated by the sets (\ref{12-18}) and
(\ref{12-19}) respectivelly. We choose a basis in the algebra octonions $\mathbb O$ such that
the subgroup $H_1\in SO(8)$ is stabilizer of identity element of $\mathbb O$. It follows from
the Moufang identity (\ref{12-21}) that
\begin{equation}\label{14-06}
R_{ab}R^{-1}_{b}R^{-1}_{a}\in H_1
\end{equation}
for any $a,b\in\mathbb S^7$. We consider a right coset $H_1g$ of $SO(8)$. Since the set $X$ in
(\ref{12-18}) generates $SO(8)$, the element
\begin{equation}\label{14-05}
g=R_{a_1}\dots R_{a_k}.
\end{equation}
Multiplying (\ref{14-05}) by suitable elements of the form (\ref{14-06}), we get the element
$R_{c}$ as a representative of $H_1g$.
\par
On the other hand, it follows from (\ref{12-09}) and (\ref{12-07}) that the product
\begin{equation}
\boldsymbol a\boldsymbol b=-(\boldsymbol a,\boldsymbol b)+\boldsymbol a\times\boldsymbol
b,\label{14-10}
\end{equation}
where $\boldsymbol a\times\boldsymbol b=\frac12[\boldsymbol a,\boldsymbol b]$. Using
properties of the algebra $\mathbb O$, we prove the equalities
\begin{equation}\label{14-22}
-(\boldsymbol b, \boldsymbol b)\boldsymbol a=(\boldsymbol a\boldsymbol b)\boldsymbol
b=-(\boldsymbol a,\boldsymbol b)\boldsymbol b-(\boldsymbol a\times\boldsymbol b, \boldsymbol
b)+(\boldsymbol a\times\boldsymbol b)\times\boldsymbol b.
\end{equation}
It follows from (\ref{14-22}) that
\begin{align}
(\boldsymbol a\times\boldsymbol b)\times\boldsymbol b&=(\boldsymbol a,\boldsymbol
b)\boldsymbol b-(\boldsymbol b,\boldsymbol b)\boldsymbol a,\label{14-07}\\
(\boldsymbol a\times\boldsymbol b,\boldsymbol b)&=0.\label{14-08}
\end{align}
Using (\ref{14-07}), we find a solution $\boldsymbol b$ of the system
\begin{equation}\label{14-09}
\begin{aligned}
\boldsymbol a\times\boldsymbol b&=\boldsymbol c,\\
-(\boldsymbol a,\boldsymbol b)&=c_0,
\end{aligned}
\end{equation}
where the vectors $\boldsymbol a$ and $\boldsymbol c$ satisfy the equalities $(\boldsymbol
a,\boldsymbol a)=1$ and $(\boldsymbol a,\boldsymbol c)=0$. This solution is
\begin{equation}
\boldsymbol b=-c_0\boldsymbol a+\boldsymbol c\times\boldsymbol a.
\end{equation}
Linearizing the identity (\ref{14-08}), we find the scalar square
\begin{equation}\label{14-11}
(\boldsymbol b,\boldsymbol b)=c_0^2+(\boldsymbol c,\boldsymbol c).
\end{equation}
Comparing  (\ref{14-09}) with (\ref{14-10}) and taking into account (\ref{14-11}), we see that
any element $c\in\mathbb S^7$ can be represented as
\begin{equation}\label{14-12}
c=\boldsymbol a\boldsymbol b.
\end{equation}
\par
By proved above, the coset $H_1g=H_1R_{c}$ for some $c\in\mathbb S^7$. We multiply $R_{c}$ by
the element
\begin{equation}
R_{\boldsymbol a}R_{\boldsymbol b}R^{-1}_{\boldsymbol a\boldsymbol b}\in H_1.
\end{equation}
Then, by (\ref{14-12}) we get the element $R_{\boldsymbol a}R_{\boldsymbol b}$ as a
representative of $H_1g$. Since this element lies in $Spin(7)$, it follows that the equality
(\ref{14-04}) is proved for $k=1$.

\subsection{The case $k\ne1$.}

We use below an explicit form of the operators $R_{e_{i}}$ in the canonical basis of $\mathbb
O$. Using the multiplication law (\ref{12-09}), we can easily find the required expressions.
We have
\begin{equation}\label{14-13}
R_{e_i}=e_{i0}+\frac12c_{ijk}e_{jk},
\end{equation}
where $e_{mn}$ are skew-symmetric $8\times8$ matrices with the elements
\begin{equation}
(e_{mn})^{\alpha}_{\beta}=\delta_{m\beta}\delta^{\alpha}_{n}
-\delta_{n\beta}\delta^{\alpha}_{m}.
\end{equation}
Since the matrices $R_{e_i}$ and $[R_{e_i},R_{e_j}]$ are linearly independent over $\mathbb
R$, they form a basis of a Lie algebra $A$ that is isomorphic to $so(8)$. Suppose
\begin{equation}\label{14-21}
I=\begin{pmatrix} E&0\\0&-E\end{pmatrix},\qquad J=\begin{pmatrix} 0&E\\-E&0\end{pmatrix},
\end{equation}
where $E$ is the identity $4\times4$ matrix. It is obvious that the transformation
\begin{equation}
R_{e_{i}}\to IR_{e_{i}}I
\end{equation}
may be extended to an involutive automorphism of $A$. With respect to this automorphism the
algebra $A$ is decomposed into the direct sum (\ref{12-20}) of proper subspaces $A^{+}$ and
$A^{-}$. Using the representation (\ref{14-13}), we prove that
\begin{equation}\label{14-16}
\begin{aligned}
IR_{e_{i}}I&=R_{e_{i}}\\
IR_{e_{i}}I&=-R_{e_{i}}
\endaligned
\qquad \aligned
&\text{for}\quad i=1,2,3,\\
&\text{for}\quad i=4,5,6,7.
\end{aligned}
\end{equation}
A simple calculation shows that $\dim A^{+}=12$ and $\dim A^{-}=16$. Therefore the
corresponding symmetric space is isomorphic to $SO(8)/H_4$.
\par
Now we consider the transformation
\begin{equation}\label{14-14}
R_{e_{i}}\to JR_{e_{i}}J^{-1}.
\end{equation}
Once again using (\ref{14-13}), we prove that
\begin{equation}\label{14-17}
\begin{aligned}
JR_{e_{i}}J^{-1}&=R_{e_{i}}\\
JR_{e_{i}}J^{-1}&=-R_{e_{i}}
\end{aligned}
\qquad \begin{aligned}
&\text{for}\quad i=1,2,4,5,6,\\
&\text{for}\quad i=3,7.
\end{aligned}
\end{equation}
Extending (\ref{14-14}) to an involutive automorphism of $A$, we get that $\dim A^{+}=16$ and
$\dim A^{-}=12$. Hence, the corresponding symmetric space is isomorphic to $SO(8)/H_2$.
\par
Finally, we consider the transformation
\begin{equation}\label{14-15}
R_{e_{i}}\to JR_{e_{i}}J.
\end{equation}
Since the transformation (\ref{14-15}) is a composition of (\ref{14-14}) and the
transformation $R_{e_{i}}\to -R_{e_{i}}$, we have the equalities
\begin{equation}\label{14-18}
\begin{aligned}
JR_{e_{i}}J&=R_{e_{i}}\\
JR_{e_{i}}J&=-R_{e_{i}}
\end{aligned}
\qquad \begin{aligned}
&\text{for}\quad i=3,7,\\
&\text{for}\quad i=1,2,4,5,6.
\end{aligned}
\end{equation}
Using (\ref{14-14}), we easily prove that the transformation (\ref{14-15}) may be extended to
an involutive automorphism of $A$. It is obvious that $\dim A^{+}=13$ and $\dim A^{-}=15$.
Therefore the corresponding symmetric space is isomorphic to $SO(8)/H_3$.
\par
We extend the involutive automorphism of $A$ defined by (\ref{14-16}), (\ref{14-17}), or
(\ref{14-18}) to an automorphism $\sigma$ of the corresponding simply connected Lie group
$Spin(8)$. It follows from (\ref{12-12}) that this group can be embedded into the Clifford
algebra $Cl_{0,7}(\mathbb R)$. Suppose $\Gamma_{i}$ is a prototype of $R_{e_i}$ relative to
the homomorphism (\ref{12-11}). It is obvious that $\Gamma_{i}\in Spin(8)$. On the other hand,
it follows from (\ref{12-10}) that the matrices $\Gamma_{i}$ generate $Cl_{0,7}(\mathbb R)$.
Hence, $\Gamma_{i}\notin Spin(7)$. Now, let $\widetilde{H}_{k}$ be a subgroup of $Spin(8)$
that is invariant under $\sigma$. Then it follows from (\ref{14-16}), (\ref{14-17}),
(\ref{14-18}) that $\Gamma_{i}\in\widetilde{H}_{k}$ for some value of $i$.
\par
Further, let the matrix $\Gamma_{i}\in\widetilde{H}_{k}$ and let $\widetilde{H}_{k}g$ be a
coset of $Spin(8)$. Since $Spin(7)$ is a maximal subgroup in $Spin(8)$, the element $g$ can be
represent by a product of $\Gamma_{i}$ and elements of $Spin(8)$. Now, note that the algebra
$\text{End}\,\mathbb O$ satisfies the identity
\begin{equation}\label{14-19}
R_{x}R_{y}R_{x}=R_{xyx},
\end{equation}
which is a direct corollary of (\ref{12-21}). Since $R_{e_{i}}R_{\bar e_{i}}=1$, it follows
that
\begin{equation}\label{14-20}
R_{e_{i}}R_{\boldsymbol a}R_{\boldsymbol b}=R_{e_{i}\boldsymbol a\bar
e_{i}}R_{e_{i}\boldsymbol b\bar e_{i}}R_{e_{i}},
\end{equation}
where we do not sum on the recurring indexes. Obviously, the products $e_{i}\boldsymbol
a\,\bar e_{i}$ and $e_{i}\boldsymbol b\,\bar e_{i}$ are vector octonions. Since a restriction
of the homomorphism (\ref{12-11}) to $Spin(7)$ is injection, it follows from (\ref{12-19}) and
(\ref{14-20}) that
\begin{equation}
\Gamma_{i}f\Gamma^{-1}_{i}\in Spin(7)
\end{equation}
for any $f\in Spin(7)$. Hence, the element $g$ can be represent in the form
$g=\Gamma_{i}^{p}f$. Since $\Gamma_{i}\in\widetilde{H}_{k}$, it follows that the element $g\in
Spin(7)$. Mapping $Spin(8)$ onto $SO(8)$, we prove the equality (\ref{14-04}) for $k\ne1$.

\section{Three and five dimensions}

In three Euclidean dimensions, the single spinor representation is two-dimensional and
pseudoreal. Therefore, the projection operator $\Omega$ may be represented as $2\times2$
Hermitian matrix. Since the group $Spin(3)\simeq SU(2)$, it follows that the matrix $\Omega$
can be reduced to the form (\ref{13-07}) by the transformation (\ref{13-08}).
\par
Now we consider five Euclidean dimensions. In these dimensions, the relevant isomorphism is
$Spin(5)\simeq Sp(2)$, which implies that the single spinor representation in four-dimensional
and pseudoreal. Hence, we must prove that
\begin{equation}\label{15-01}
B^{-1}\Omega B=\tilde\Omega=\text{diag}\{1,\dots,1,0,\dots,0\}.
\end{equation}
for some $B\in Sp(2)$. Since the space of spinor representation of $Spin(5)$ is a
four-dimensional unitary space, the Hermitian matrix $\Omega$ can be reduced to the form
(\ref{13-07}) by the transformation
\begin{equation}\label{15-02}
\tilde\Omega=U\Omega U^{-1},
\end{equation}
where $U\in SU(4)$. As above, it follows from (\ref{15-01}) and (\ref{15-02}) that
\begin{equation}\label{15-03}
\tilde\Omega\tilde B=\tilde B\tilde\Omega,
\end{equation}
where the matrix $\tilde B=UB$. The general solution of the Eq. (\ref{15-03}) has the form
\begin{equation}
\tilde B=\begin{pmatrix}\tilde B_1&0\\0&\tilde B_2\end{pmatrix},
\end{equation}
where $\tilde B_1$ and $\tilde B_2$ are unitary matrices such that $\det\tilde B=1$. It is
obvious that
\begin{equation}
\tilde B\in H_{k}\simeq S(U(k)\times U(4-k)),\qquad 1\leqslant k\leqslant2.
\end{equation}
Since $U=\tilde BB^{-1}$, the equality (\ref{15-01}) is true if the group
\begin{equation}\label{15-04}
SU(4)=H_{k}Sp(2),
\end{equation}
i.e. if any element $SU(4)$ can be represented as the product $g=hf$, where $h\in H_{k}$ and
$f\in Sp(2)$. We will prove the equality (\ref{15-04}).

\subsection{The case $k=1$}

As before, we will use properties of the algebra $\mathbb O$. We fix first the field $\mathbb
C$ in $\mathbb O$ by the condition $e_1\in\mathbb C$. Further, any two elements of $\mathbb O$
generate an associative subalgebra. Therefore,
\begin{equation}
x(yz)=(xy)z
\end{equation}
for any $x,y\in\mathbb C$ and $z\in\mathbb O$. It follows that we may consider $\mathbb O$ as
a (left) vector space over $\mathbb C$ relative to the multiplication $xz$, where $x\in\mathbb
C$ and $z\in\mathbb O$. Obviously, $\mathbb O$ is four dimensional over $\mathbb C$. For
$x,y\in\mathbb O$ we define
\begin{equation}\label{15-07}
\langle x,y\rangle=(x,y)-e_1(e_1x,y).
\end{equation}
Then $\langle x,y\rangle\in\mathbb C$. Using the identities (\ref{12-13}) and (\ref{12-14}),
we prove the equalities
\begin{equation}\label{15-06}
\langle e_1x,y\rangle=e_1\langle x,y\rangle=-\langle x,e_1y\rangle.
\end{equation}
Hence, $\langle x,y\rangle$ is a Hermitian form in $\mathbb O$ over $\mathbb C$. If $\langle
x,y\rangle=0$, then $(x,y)=0$, since $1$ and $e_1$ are independent over $\mathbb R$. Since the
form (\ref{12-07}) is positive definite, it follows that the Hermitian form (\ref{15-06}) is
nondegenerate.
\par
Further, let $V$ be a linear  span of the elements $1,e_1,e_2$. Denote by $\mathbb C^{\perp}$
and $V^{\perp}$ the orthogonal complements to $\mathbb C$ and $V$ in $\mathbb O$ and define
the sets
\begin{align}
\mathbb S^5&=\{\boldsymbol a\in\mathbb C^{\perp}\mid n(\boldsymbol a)=1\},\\
\mathbb S^4&=\{\boldsymbol a\in V^{\perp}\mid n(\boldsymbol a)=1\}.
\end{align}
Now, note that the elements $\Gamma_2,\Gamma_3,\dots,\Gamma_7$ of the Clifford algebra
$Cl_{0,7}(\mathbb R)$ generate the subalgebra $Cl_{0,6}(\mathbb R)$. It follows from Table 1
that $Cl_{0,6}(\mathbb R)$ is isomorphic to the simple matrix algebra $\mathbb R(8)$.
Therefore, the restriction of the homomorphism (\ref{12-11}) to $Cl_{0,6}(\mathbb R)$ is
injection. It is obvious that the restriction of this homomorphism to the algebra
$Cl_{0,5}(\mathbb R)$ with the generators $\Gamma_3,\dots,\Gamma_7$ is also injection. Hence
the sets
\begin{align}
Z_1&=\{R_{\boldsymbol a}R_{\boldsymbol b}\mid\boldsymbol a,\boldsymbol b\in\mathbb S^5\},
\label{15-08}\\
Z_2&=\{R_{\boldsymbol a}R_{\boldsymbol b}\mid\boldsymbol a,\boldsymbol b\in\mathbb
S^4\}\label{15-09}
\end{align}
generate the groups $G_1$ and $G_2$, which are isomorphic to $Spin(6)$ and $Spin(5)$,
respectivelly. Further, if follows from (\ref{12-09}) that the elements $1,e_2,e_4,e_6$ form a
basis of $\mathbb O$ over $\mathbb C$. We will prove that in this basis the groups $G_1$ and
$G_2$ coincide with $SU(4)$ and $Sp(2)$. Indeed, for all $x\in\mathbb O$ and $\boldsymbol
a,\boldsymbol b\in\mathbb S^5$ the equality
\begin{equation}\label{15-05}
(xR_{\boldsymbol a}R_{\boldsymbol b})e_1=(xe_1)R_{\boldsymbol a}R_{\boldsymbol b}
\end{equation}
is true. This equality can easily obtain with the help of the multiplication law
(\ref{12-09}). Using (\ref{15-05}) and (\ref{15-06}), we prove that the form (\ref{15-07}) is
invariant under elements of (\ref{15-08}). Therefore, elements of $G_1$ may be represented as
$4\times4$ unitary matrices. Our assertion follows then from the isomorphisms $Spin(6)\simeq
SU(4)$ and $Spin(5)\simeq Sp(2)$. In addition, we note that
\begin{equation}\label{15-11}
H=\{g\in G_1\mid 1g=1\}
\end{equation}
is a group that isomorphic to $SU(3)$.
\par
Now suppose $Hg$ is the right coset of $G_1$, where $H$ is defined by (\ref{15-11}).
Obviously, the element
\begin{equation}\label{15-16}
g_1=R_{e_2}R_{e_4}
\end{equation}
belong to $G_1$ but do not belong to $G_2$. On the other hand, the groups $SU(4)$ and $Sp(2)$
are the double cover of $SO(6)$ and $SO(5)$, respectivelly. Therefore, $G_2$ is maximal the
subgroup of $G_1$. Hence $g$ can be represent as a product of elements of $G_2\cup\{g_1\}$.
Using (\ref{14-19}) and (\ref{12-09}), we prove  that
\begin{equation}\label{15-10}
R_{e_2}R_{\boldsymbol a}=R_{\bar{\boldsymbol a}}R_{e_2}
\end{equation}
for all $\boldsymbol a\in S^4$. Since $G_2$ is generated by (\ref{15-09}) and
${\boldsymbol{\bar a}\in S^4}$, it follows that
\begin{equation}\label{15-14}
g=(R_{e_2}R_{\boldsymbol b})^{\sigma}f,\qquad\sigma\in\{0,1\},
\end{equation}
where $\boldsymbol b\in\mathbb S^4$ and $f\in G_2$. If $\sigma=0$, then we can choose an
element of $G_2$ as a representative of $Hg$.
\par
Let $\sigma=1$. Since the product $R_{\bar e_4}R_{\boldsymbol b}\in G_2$, it follows that
\begin{equation}\label{15-15}
g=g_1f',
\end{equation}
where $f'\in G_2$. Suppose
\begin{equation}
h=R_{e_5}R_{e_3}R_{e_4}R_{e_2}.
\end{equation}
It follows from (\ref{12-09}) that $1h=1$. Hence $h$ belongs to the subgroup (\ref{15-11}).
Therefore
\begin{equation}
Hg=Hhg_1f'=Hf'',
\end{equation}
where again $f''\in G_2$. Thus, we can choose a representative of $Hg$ in the subgroup $G_2$.
The equality (\ref{15-04}) is proven for $k=1$.

\subsection{The case $k=2$}

Obviously, the matrices $R_{ij}=\frac12[R_{e_i},R_{e_j}]$ are independent over $\mathbb R$. In
addition, it is follows from (\ref{12-10}) that they satisfy the following commutation
relations:
\begin{equation}
[R_{ij},R_{kl}]=\delta_{ik}R_{jl}+\delta_{jl}R_{ik}-\delta_{il}R_{jk}-\delta_{jk}R_{il}.
\end{equation}
Hence, the matrices $R_{ij}$ form a basis of the algebra $A\simeq so(7)$. We consider the
transformation
\begin{equation}\label{15-12}
R_{e_{i}}\to J(KR_{e_{i}}K)J,
\end{equation}
where the matrix $J$ is defined in (\ref{14-21}) and the matrix
\begin{equation}
K=\text{diag}(1,-1,-1,1,-1,1,1,-1).
\end{equation}
Using the explicit form (\ref{14-13}) of $R_{e_{i}}$, we prove that
\begin{equation}\label{15-13}
\begin{aligned}
J(KR_{e_{i}}K)J&=R_{e_{i}}\\
J(KR_{e_{i}}K)J&=-R_{e_{i}}
\end{aligned}
\qquad\begin{aligned}
&\text{for}\quad i=1,2,3,4,\\
&\text{for}\quad i=5,6,7.
\end{aligned}
\end{equation}
Obviously, the transformation (\ref{15-12}) can be extend to an involutive automorphism of
$A$. We consider the subalgebra $A_1\subset A$ generated by the elements $R_{ij}$, where
$i,j=2,\dots,7$. It is obvious that $A_1\simeq so(6)$. With respect to this automorphism the
algebra $A_1$ can be decomposable into the direct sum (\ref{12-20}) of the proper subspaces
$A_1^{+}$ and $A_1^{-}$. It follows from (\ref{15-13}) that $A_1^{+}\simeq so(3)\oplus so(3)$.
Since $so(6)\simeq su(4)$ and $so(3)\simeq su(2)$, it follows that the corresponding symmetric
space is isomorphic to $SU(4)/H_2$.
\par
We extend the involutive automorphism of $A$ defined by (\ref{15-12}) to an automorphism
$\tilde\sigma$ of the corresponding simply connected Lie group $Spin(7)$. We suppose that this
group is embedded into the Clifford algebra $Cl_{0,7}(\mathbb R)$. Since the restriction of
the homomorphism (\ref{12-11}) to $Spin(7)$ is injection, $\tilde\sigma$ induces an involutive
automorphism $\sigma$ of $G\in\text{Aut}\,\mathbb O$. It is obvious that $G\simeq Spin(7)$. On
the other hand, for all $\boldsymbol a,\boldsymbol b\in\mathbb S^6$ the product
\begin{equation}
R_{\boldsymbol a}R_{\boldsymbol b}=-R_{(\boldsymbol a,\boldsymbol b)}+\frac12[R_{\boldsymbol
a},R_{\boldsymbol b}].
\end{equation}
Using (\ref{12-19}), we prove that the automorphism $\sigma$ of $G$ is defined by
(\ref{15-13}). Obviously, the restrictions of $\sigma$ to $G_1$ and $G_2$ can be also defined
by (\ref{15-13}).
\par
Now suppose $H$ is a subgroup of $G_1$ invariant under the automorphism $\sigma$, and $Hg$ is
a right coset of $G_1$. As in the arguments above, we represent $g$ in the form (\ref{15-14}).
If $\sigma=0$, then we can choose an element of $G_2$ as a representative of $Hg$. If
$\sigma=1$, then $g$ has the form (\ref{15-15}). But it follows from (\ref{15-13}) that the
element (\ref{15-16}) is invariant under the automorphism $\sigma$. Therefore it belongs to
$H$. Hence we can choose a representative of $Hg$ in the subgroup $G_2$. Since the groups $H$
and $H_2$ are isomorphic, it follows that the equality (\ref{15-04}) is proved for $k=2$. This
completes the proof of Theorem 1.

\section{Classification of BPS equations}

We have proved that to within equivalence there exists a unique system of BPS equations for
every pair of values $d\leq 8$ and $\nu=\nu(d)$. In this section, we find all such systems of
equations. However, we present first a general method allowing to obtain the systems of BPS
equations.
\par
Let $V$ be a space of irreducible spinor representation of $Spin(d)$ and
$\Omega_1,\dots,\Omega_{2^{s}}:V\to V$ be a finite set of linear operators satisfying the
conditions
\begin{equation}\label{16-01}
\sum^{2^{s}}_{\alpha=1}\Omega_{\alpha}=1,\qquad
\Omega_{\alpha}\Omega_{\beta}=\delta_{\alpha\beta}\Omega_{\beta}.
\end{equation}
We say that the operators $\Omega_1,\dots,\Omega_{2^{s}}$ make up a total orthogonal system of
idempotent operators and the corresponding matrices make up a total orthogonal system of
idempotent matrices. Obviously, every such operator is a projector onto a subspace in $V$.
Moreover, with respect to this system of projectors the space $V$ decomposes into the direct
sum
\begin{equation}
V=V_1\oplus\dots\oplus V_{2^{s}}
\end{equation}
of the subspaces $V_{\alpha}=\text{Im}\,\Omega_{\alpha}$. The idempotent $\Omega_{\alpha}$ is
called primitive if it is not a sum of two nonzero mutually orthogonal idempotents. It is
obvious that any projector is a sum of mutually orthogonal idempotents. Finally, if every
idempotent in (\ref{16-01}) is primitive, than we have a total orthogonal system of primitive
idempotents.
\par
Since irreducible spinor representations of $Spin(d)$ are realized in the algebra
$Cl_{0,d-1}(\mathbb R)$, we will find a total orthogonal system of primitive idempotents in
this algebra. To this end, we choose a subset of monomials $E_1,\dots,E_{s}$ in (\ref{12-03})
such that
\begin{equation}\label{16-04}
E_{i}^2=1,\qquad [E_{i},E_{j}]=0.
\end{equation}
Further, we impose the condition (\ref{12-22}) on the gamma matrices and define the $2^{s}$
matrices
\begin{equation}\label{16-02}
\Omega[\alpha_1,\dots,\alpha_{s}]=\frac{1}{2^{s}}\prod^{s}_{i=1}(1+\alpha_{i}E_{i}),
\end{equation}
where $\alpha_{i}=\pm1$. It is easily shown that these matrices are Hermitian and satisfy the
equalities (\ref{16-01}). Since such notations of matrices are some few inconveniently, we
introduce new notations. To this end, we denote the matrices (\ref{16-02}) by
\begin{equation}
\Omega_1=\Omega[1,\dots,1],\quad\Omega_2=\Omega[1,\dots,-1],\quad\dots,\quad\Omega_{2^{s}}
=\Omega[-1,\dots,-1].
\end{equation}
Notice that this way of ranking is used in the binary number system. Besides, we suppose that
\begin{equation}
\Omega_{\alpha_1\dots\alpha_{r}}=\sum^{r}_{i=1}\Omega_{\alpha_{i}}.
\end{equation}
\par
Further, with respect to the system of orthogonal idempotents (\ref{16-02}) the algebra
$Cl_{0,d-1}(\mathbb R)$ decomposes into the direct sum
\begin{equation}\label{16-03}
Cl_{0,d-1}(\mathbb R)=I_{1}+\dots+I_{2^{s}}
\end{equation}
of left ideals $I_{\alpha}=Cl_{0,d-1}(\mathbb R)\Omega_{\alpha}$. And also, the idempotent
$\Omega_{\alpha}$ is primitive if and only if the left ideal $I_{\alpha}$ is minimal. It
follows from Table 1 that all minimal left ideals of $Cl_{0,d-1}(\mathbb R)$ are isomorphic.
Obviously, dimensions of minimal left ideals in $Cl_{0,d-1}(\mathbb R)$ and irreducible spinor
representations of $Spin(d)$ coincide. Let this dimension over $\mathbb R$ be $2^{p}$. Then
the quantity of mutually orthogonal primitive idempotents is $2^{d-p}$. Hence,
$Cl_{0,d-1}(\mathbb R)$ contains always $s=d-p$ monomials $E_1,\dots,E_{s}$ satisfying the
conditions (\ref{16-04}).
\par
After we find the primitive idempotent (\ref{16-02}) (or monomials $E_{i}$) in алгебре
$Cl_{0,d-1}(\mathbb R)$, we must find its isomorphic images in $Cl_{d,0}(\mathbb R)$. We can
easy do it if we write the isomorphism
\begin{equation}
Cl_{0,d-1}(\mathbb R)\to Cl_{d,0}^0(\mathbb R)
\end{equation}
in the explicit form
\begin{equation}\label{16-05}
\Gamma_{a_1\dots a_{k}}\to \left\{ \aligned
(\Gamma_{a_1\dots a_{k}})^{\dag}\qquad &\text{for}\,\,\text{even}\,\,k,\\
(\Gamma_{a_1\dots a_{k}}\Gamma_{d})^{\dag}\qquad &\text{for}\,\,\text{odd}\,\,k.
\endaligned\right.
\end{equation}
Having the total orthogonal system of primitive idempotents in $Cl_{d,0}^0(\mathbb R)$, we
easy find the BPS equations from (\ref{13-02}). Note that the fraction $\nu$ of the unbroken
supersymmetry can be found as
\begin{equation}\label{16-06}
\nu=\frac{\dim I}{\dim Cl_{0,d-1}(\mathbb R)},
\end{equation}
where $I$ is a left ideal of $Cl_{0,d-1}(\mathbb R)$ corresponding to the idempotent $\Omega$.
The dimension of $I$ can be found in Table 1. Now, we will construct BPS equations in the
concrete dimensions.

\subsection{The dimension $d\leq3$}

In these dimensions, the algebra $Cl_{0,d-1}(\mathbb R)$ is a division algebra. Therefore, any
its left ideal is either trivial or coinciding with $Cl_{0,d-1}(\mathbb R)$. It follows that
the idempotent $\Omega=0$ or 1. Thus, any system of BPS equations has only the trivial
solution $F_{ab}=0$.

\subsection{Four dimensions}

The algebra $Cl_{0,3}(\mathbb R)$ decomposes into the direct sum of two minimal left ideals.
Using the decomposition (\ref{16-03}), we find $s=1$. Further, we choose the monomial
$E_1=\Gamma_{123}$ in $Cl_{0,3}(\mathbb R)$. Obviously, the square $E_1^2=1$. Using the
mapping (\ref{16-05}), we find the image of $E_1$ in $Cl_{4,0}^0(\mathbb R)$ and next
construct the total orthogonal system of primitive idempotents
\begin{equation}\label{16-07}
\Omega_{\alpha}=\frac12(1\pm\Gamma_{1234}),
\end{equation}
where $\alpha=1,2$. Substituting $\Omega_1$ in Eq. (\ref{13-02}), we get the BPS equations
\begin{equation}\label{16-25}
F_{ab}=\frac12\varepsilon_{abcd}F_{cd},
\end{equation}
where $\varepsilon_{abcd}$ is the completely antisymmetric identity four tensor. Using
(\ref{16-06}), we find $\nu=1/2$. Note that we consider the chiral representation.

\subsection{Five dimensions}

The algebra $Cl_{0,4}(\mathbb R)$ also decomposes into the direct sum of two minimal left
ideals. Hence, $s=1$. We choose the monomial $E_1=\Gamma_{1234}$ in $Cl_{0,5}(\mathbb R)$,
find its image in $Cl_{5,0}^0(\mathbb R)$, and construct the total orthogonal system of
primitive idempotents. Obviously, this system coincides with (\ref{16-07}). Substituting
$\Omega_1$ in (\ref{13-02}), we get the BPS equations
\begin{equation}
\begin{aligned}
F_{ab}&=\frac12\varepsilon_{abcd}F_{cd},\\
F_{a5}&=0.
\end{aligned}
\end{equation}
It is obvious that the fraction of the unbroken supersymmetry $\nu=1/2$.

\subsection{Six dimensions}

The algebra $Cl_{0,5}(\mathbb R)$ decomposes into the direct sum of four minimal left ideals.
In this case, $\nu=1/4$ and $s=2$. We choose the monomials $E_1=\Gamma_{125}$ and
$E_2=\Gamma_{345}$ in this algebra. Obviously, they satisfy the conditions (\ref{16-04}).
Using (\ref{16-05}), we find images of these monomials in $Cl_{6,0}^0(\mathbb R)$ and
construct the total orthogonal system of primitive idempotents
\begin{equation}\label{16-09}
\Omega_{\alpha}=\frac14(1\pm\Gamma_{1234})(1\pm\Gamma_{1256}).
\end{equation}
Substituting $\Omega_1$ in Eq. (\ref{13-02}), we get the following BPS equations:
\begin{equation}\label{16-10}
\begin{gathered}
F_{12}+F_{43}+F_{65}=0,\\
\vspace{1\jot} \begin{aligned}
F_{13}+F_{24}&=0,\\
F_{14}+F_{32}&=0,\\
F_{15}+F_{26}&=0,
\end{aligned}\qquad
\begin{aligned}
F_{16}+F_{52}&=0,\\
F_{35}+F_{64}&=0,\\
F_{36}+F_{45}&=0.
\end{aligned}
\end{gathered}
\end{equation}
Now we consider the sum $\Omega_{12}$ of two primitive idempotents $\Omega_1$ and $\Omega_2$
\begin{equation}\label{16-08}
\Omega_{12}=\frac12(1+\Gamma_{1234}).
\end{equation}
Obviously, the prototype of $\Omega_{12}$  in $Cl_{0,5}(\mathbb R)$ is the identity of a left
ideal $I$. Since $\dim I=16$, it follows that $\nu=2/4$. Substituting (\ref{16-08}) in
(\ref{13-02}), we find the BPS equations
\begin{equation}
\begin{aligned}
F_{ab}&=\frac12\varepsilon_{abcd}F_{cd},\\
F_{a5}&=F_{a6}=0.
\end{aligned}
\end{equation}
If we calculate the sum $\Omega_{123}$ of three primitive idempotents of the form
(\ref{16-09}) and substitute it to (\ref{13-02}), then we get the system of BPS equations
having only trivial solution. The alternative way to get this system is following. We find
systems of the form (\ref{16-10}) for every $\Omega_{\alpha}$ ($\alpha=1,2,3$). Such systems
is called primitive. Then the system corresponding to $\Omega_{123}$ is a system joining the
systems for every $\Omega_{\alpha}$. It can be easily be checked that this joined system has
only trivial solution.

\subsection{Seven dimensions}

The algebra $Cl_{0,6}(\mathbb R)$ decomposes into the direct sum of eight minimal left ideals.
In this case, $\nu=1/8$ and $s=3$. We choose the monomials $E_1=\Gamma_{1234}$,
$E_2=\Gamma_{1256}$, and $E_3=\Gamma_{164}$ in $Cl_{0,6}(\mathbb R)$. Using (\ref{16-05}), we
find images of the monomials in $Cl_{7,0}^0(\mathbb R)$ and construct the total orthogonal
system of primitive idempotents
\begin{equation}\label{16-13}
\Omega_{\alpha}=\frac18(1\pm\Gamma_{1234})(1\pm\Gamma_{1256})(1\pm\Gamma_{1476}).
\end{equation}
Substituting $\Omega_1$ in (\ref{13-02}), we get the following BPS equations
\begin{equation}\label{16-24}
\begin{gathered} F_{12}+F_{43}+F_{65}=0,\\
\vspace{1\jot} \begin{aligned}
F_{13}+F_{24}+F_{75}&=0,\\
F_{14}+F_{32}+F_{67}&=0,\\
F_{15}+F_{26}+F_{37}&=0,
\end{aligned}\qquad
\begin{aligned}
F_{16}+F_{52}+F_{74}&=0,\\
F_{17}+F_{53}+F_{46}&=0,\\
F_{27}+F_{54}+F_{63}&=0.
\end{aligned}
\end{gathered}
\end{equation}
Further, we consider the sum $\Omega_{12}$ of two primitive idempotents $\Omega_1$ and
$\Omega_2$
\begin{equation}\label{16-11}
\Omega_{12}=\frac14(1+\Gamma_{1234})(1+\Gamma_{1256}).
\end{equation}
The dimension of left ideal corresponding to $\Omega_{12}$ is 16. Therefore, $\nu=2/8$. The
corresponding system of BPS equations has the form
\begin{equation}\label{16-12}
\begin{aligned}
F_{12}+F_{43}+F_{65}&=0,\\
F_{13}+F_{24}&=0,\\
F_{14}+F_{32}&=0,\\
F_{15}+F_{26}&=0,\\
\end{aligned}\qquad
\begin{aligned} F_{a7}&=0,\\
F_{16}+F_{52}&=0,\\
F_{35}+F_{64}&=0,\\
F_{36}+F_{45}&=0.
\end{aligned}
\end{equation}
Now we find the BPS equations corresponding to $\Omega_{123}$. To this end, we write BPS
equations for
\begin{equation}
\Omega_{3}=\frac18(1+\Gamma_{1234})(1-\Gamma_{1256})(1+\Gamma_{1476}).
\end{equation}
and join them with the system (\ref{16-12}). As result, we get the following BPS equations:
\begin{equation}
\begin{aligned}
F_{ab}&=\frac12\varepsilon_{abcd}F_{cd},\\
F_{a5}&=F_{a6}=F_{a7}=0.
\end{aligned}
\end{equation}
Since the dimension of the corresponding left ideal is $24$, it follows that $\nu=3/8$. It can
be easily be checked that the system of BPS equations constructed by means of four primitive
idempotents has only trivial solution.

\subsection{Eight dimensions}

The algebra $Cl_{0,7}(\mathbb R)$ decomposes into the direct sum of 16 minimal left ideals.
Hence, $s=4$. We choose in $Cl_{0,6}(\mathbb R)$ the monomials $E_1=\Gamma_{1234}$,
$E_2=\Gamma_{1256}$, $E_3=\Gamma_{1476}$, and the monomial $E_4=\Gamma_{*}$ defined in
(\ref{13-10}). We find its images in $Cl_{8,0}^0(\mathbb R)$ and construct the total
orthogonal system of primitive idempotents
\begin{equation}\label{16-18}
\Omega_{\alpha}=\frac{1}{16}(1\pm\Gamma_{*})(1\pm\Gamma_{1234})(1\pm\Gamma_{1256})
(1\pm\Gamma_{1476}).
\end{equation}
Obviously, we can find the BPS systems by the method that was used above. However, all these
systems had been found in~\cite{bak02}. Therefore we simply list them.

\medskip\noindent
(1) $\nu=1/16$,\, $\Omega=\Omega_1$

\begin{equation}\label{16-19}
\begin{aligned}
F_{12}+F_{43}+F_{65}+F_{78}&=0,\\
F_{13}+F_{24}+F_{75}+F_{86}&=0,\\
F_{14}+F_{32}+F_{67}+F_{85}&=0,\\
F_{15}+F_{26}+F_{37}+F_{48}&=0,\\
F_{16}+F_{52}+F_{74}+F_{38}&=0,\\
F_{17}+F_{53}+F_{46}+F_{82}&=0,\\
F_{18}+F_{27}+F_{54}+F_{63}&=0.
\end{aligned}
\end{equation}

\medskip\noindent
(2) $\nu=2/16$,\, $\Omega=\Omega_{12}$

\begin{equation}\label{16-14}
\begin{gathered} F_{12}+F_{43}+F_{65}+F_{78}=0,\\
\vspace{1\jot} \begin{aligned}
F_{13}+F_{24}&=0,\\
F_{14}+F_{32}&=0,\\
F_{15}+F_{26}&=0,\\
F_{16}+F_{52}&=0,
\end{aligned}\qquad
\begin{aligned}
F_{17}+F_{82}&=0,\\
F_{18}+F_{27}&=0,\\
F_{75}+F_{86}&=0,\\
F_{67}+F_{85}&=0,
\end{aligned}\qquad
\begin{aligned}
F_{37}+F_{48}&=0,\\
F_{38}+F_{74}&=0,\\
F_{46}+F_{53}&=0,\\
F_{54}+F_{63}&=0.
\end{aligned}
\end{gathered}
\end{equation}

\medskip\noindent
(3) $\nu=3/16$,\, $\Omega=\Omega_{123}$

\begin{equation}\label{16-15}
\begin{gathered} \begin{aligned}
F_{12}+F_{43}&=0,\\
F_{56}+F_{87}&=0,
\end{aligned}\qquad
\begin{aligned}
F_{13}+F_{24}&=0,\\
F_{57}+F_{68}&=0,
\end{aligned}\qquad
\begin{aligned}
F_{14}+F_{32}&=0,\\
F_{58}+F_{76}&=0,
\end{aligned}\\
\vspace{1\jot}\begin{aligned}
F_{15}=F_{37}=F_{62}=F_{84}&,\\
F_{16}=F_{25}=F_{38}=F_{47}&,\\
F_{17}=F_{28}=F_{53}=F_{64}&,\\
F_{18}=F_{45}=F_{63}=F_{72}&.
\end{aligned}
\end{gathered}
\end{equation}

\medskip\noindent
(4) $\nu=4/16$,\, $\Omega=\Omega_{1234}$

\begin{equation}\label{16-16}
\begin{gathered} \begin{aligned}
F_{12}+F_{43}&=0,\\
F_{56}+F_{87}&=0,
\end{aligned}\qquad
\begin{aligned}
F_{13}+F_{24}&=0,\\
F_{57}+F_{68}&=0,
\end{aligned}\qquad
\begin{aligned}
F_{14}+F_{32}&=0,\\
F_{58}+F_{76}&=0,
\end{aligned}\\
\vspace{1\jot} F_{ab}=0\quad\text{для}\quad a\in\{1,2,3,4\},\quad b\in\{5,6,7,8\}.
\end{gathered}
\end{equation}

\medskip\noindent
(5) $\nu=5/16$,\, $\Omega=\Omega_{12345}$

\begin{equation}\label{16-17}
\begin{gathered} \begin{aligned}
F_{12}=F_{34}=F_{56}=F_{78}&,\\
F_{13}=F_{42}=F_{68}=F_{75}&,\\
F_{14}=F_{23}=F_{76}=F_{85}&,
\end{aligned}\\
\vspace{1\jot} F_{ab}=0\quad\text{для}\quad a\in\{1,2,3,4\},\quad b\in\{5,6,7,8\}.
\end{gathered}
\end{equation}

\medskip\noindent
(6) $\nu=6/16$,\, $\Omega=\Omega_{123456}$

\begin{equation}
F_{12}=F_{34}=F_{56}=F_{78},
\end{equation}
and other components are zero. The system of BPS equations constructed by means of seven
primitive idempotents of the form (\ref{16-18}) has only trivial solution.

\section{Discussions and Comments}

In this paper, we systematically classified all possible BPS equations in Euclidean dimension
$d\leq8$ and presented a general method allowing to obtain the BPS equations in any dimension.
In this section, we discuss symmetries of BPS equations and their connection with the
self-dual Yang-Mills equations. Further, we find all BPS equations in the Minkowski space of
dimension $d\leq6$. In addition, we apply the obtained results to the supersymmetric
Yang-Mills theories and to the low-energy effective theory of the heterotic string.

\subsection{Symmetries of BPS equations}

First, we consider a connection between BPS states and instantons in the Euclidean Yang-Mills
theory. We note that the primitive system (\ref{16-19}) can be rewritten in the form
\begin{equation}\label{17-03}
F_{ab}=\frac12 f_{abcd}F_{cd},
\end{equation}
where $f_{abcd}$ is a completely antisymmetric tensor with the following nonzero components:
\begin{equation}\label{17-04}
\begin{aligned}
f_{1234}&=f_{1256}=f_{1357}=f_{1476}=f_{2367}=f_{2457}=f_{3465}=1,\\
f_{5678}&=f_{3476}=f_{2468}=f_{3258}=f_{1458}=f_{1368}=f_{1728}=1.
\end{aligned}
\end{equation}
Let $d<8$. Suppose that the components (\ref{17-04}) with the indices $i>d$ equal to zero.
Then we get the primitive system of BPS equations in dimension $d$. Obviously, this system has
the form (\ref{17-03}). Since any system of BPS equations is a system joining primitive
systems, it also has the form (\ref{17-03}). Thus, any BPS equation in Euclidean space of
dimension $d\leq8$ is equivalent to a self-dual Yang-Mills equation. It follows that any
solution of BPS equations in the Euclidean super Yang-Mills theory in this dimension is an
instanton solution.
\par
We consider symmetries of the BPS equations. In Euclidean dimension $d\leq8$, the group $G$ of
symmetries of BPS equations is a subgroup of $SO(8)$. On the other hand, the corresponding
projection operator $\Omega$ is invariant under this subgroup. Using the canonical form of
$\Omega$, we easily find the group $G$. We list all such group in the next table.
\par
\bigskip
{\small \noindent Table 2. Groups of symmetries of BPS equations
$$
\arraycolsep=2.2mm
\begin{array}{|c|c|c|c|c|}
\hline
d=4,5&\nu=1/2&&&\\
\hline
&SO(4)&&&\\
\hline
d=6&\nu=1/4&2/4&&\\
\hline
&SU(3)\times U(1)/Z_3&SO(4)\times SO(2)&&\\
\hline
d=7&\nu=1/8&2/8&3/8&\\
\hline
&G_2&SU(3)\times U(1)/Z_3&SO(4)\times SO(3)&\\
\hline
d=8&\nu=1/16&2/16,\,\,6/16&3/16,\,\,5/16&4/16\\
\hline
&Spin(7)&SU(4)\times U(1)/Z_4&Sp(2)\times SU(2)/Z_2&SO(4)\times SO(4)\\
\hline
\end{array}
$$}
\par\noindent
Note that these groups was first interpreted as groups of symmetries of the self-dual
Yang-Mills equations in~\cite{corr83,ward84}. In the same place, an example of self-dual
equations that differ from the BPS equations was found. These equations can be obtained if we
deduce the equality of each term in each row of (\ref{16-19}), i.e.
$F_{12}=F_{43}=F_{65}=F_{78}$, etc., a set of 21 equations. It follows that solution of the
self-dual Yang-Mills equations is not necessarily a solution of BPS equations in the Euclidean
super Yang-Mills theory.
\par
Let us discuss a possibility of generalization of Theorem 1. First note that we can consider
arbitrary spinor representations of $Spin(d)$. Then new systems of BPS equations appear in
eight dimensions. We can easily construct such system using the sum $\Omega_{+}+\Omega_{-}$ of
the idempotents
\begin{equation}
\Omega_{\pm}=\frac{1}{16}(1\pm\Gamma_{*})(1+\Gamma_{1234})(1+\Gamma_{1256}) (1+\Gamma_{1476}).
\end{equation}
Obviously, it is a system joining the system (\ref{16-24}) with the conditions $F_{a8}=0$.
Conversely, new BPS equations do not appear in dimension $d<8$. This assertion is obvious for
odd $d$, because any spinor representation of $Spin(d)$ is irreducible in such dimension. In
order that to prove this assertion for even $d<8$, we use~\cite{bak02}. In this work, all BPS
equation in even dimension $d\leq8$ was found. And also in Euclidean dimension $d<8$,
arbitrary spinor representations of $Spin(d)$ was considered. It was proved that only trivial
BPS equations are in two dimensions. In four dimensions, the chiral BPS Eq. (\ref{16-25}) and
their the antichiral analog were found. In six dimension, it was proved that BPS equations
either have the form
\begin{equation}\label{17-01}
\begin{gathered}
F_{12}+\alpha_2F_{34}+\alpha_1F_{56}=0,\\
\vspace{1\jot} \begin{aligned}
F_{13}+\alpha_2F_{42}&=0,\\
F_{15}+\alpha_1F_{62}&=0,\\
F_{35}+\alpha_1\alpha_2F_{64}&=0,
\end{aligned}\qquad
\begin{aligned}
F_{14}+\alpha_2F_{23}&=0,\\
F_{16}+\alpha_1F_{25}&=0,\\
F_{36}+\alpha_1\alpha_2F_{45}&=0,
\end{aligned}
\end{gathered}
\end{equation}
where $\alpha_1$, $\alpha_2$ are two independent signs $\pm1$, or are a corollary of
(\ref{17-01}). The problem of equivalence is not being considered in this work. Nevertheless,
we can prove that the four system (\ref{17-01}) defined by the choice of values of $\alpha_1$
and $\alpha_2$ are equivalent. Indeed, the permutation $(13)(24)$, $(15)(26)$, $(35)(46)$ of
indices of $F_{ab}$ leave invariant two system and transpose the other two with each other. In
turn, such transformations of BPS equations can be obtained by the transformations
(\ref{13-04}). Since the considered supersymmetry is global, it follows that these four
systems of BPS equations are equivalent. It is obvious also that they are equivalent to the
system (\ref{16-10}). It is sufficient to put $\alpha_1=\alpha_2=1$ in (\ref{17-01}) and then
use the permutation $(34)(56)$ of indices of $F_{ab}$. Thus, Theorem 1 is true for any spinor
representations of $Spin(d)$ in dimension $d\leq6$. Also, we prove that to within equivalence
all BPS equations found in~\cite{bak02} are the self-dual Yang-Mills equations.

\subsection{BPS equations in the Minkowski spaces}

The second possibility of generalization of Theorem 1 is connected with an investigation of
BPS equations in the Minkowski space. These equations also may be obtained by a dimensional
reduction of the $D=10$ $N=1$ super Yang-Mills theory. Note that the method used above may be
applied in this case. In particular, all constructions of Sec. 2 are remained true if we are
restricted to the dimension $d<8$. It is clear that we must correctly place the tensor indices
in the text and also use the groups $Spin(d-1,1)$, $SO(d-1,1)$ and the anti-Hermitian matrices
$i\Gamma_{d}$ instead of the groups $Spin(d)$, $SO(d)$ and the Hermitian matrices
$\Gamma_{d}$. The following weakened analog of Theorem 1 is true.
\begin{theorem}
In the Minkowski space of dimension $d\leq6$, there exists unique to within equivalence
nontrivial system of BPS equations connected with constant chiral spinor.
\end{theorem}
Indeed, the matrix $\Omega_{+}=0$ or $1$ in dimension $d=1+1$. We consider dimension $d=3+1$.
Since the group $Spin(3,1)$ is isomorphic to $Sl(2,\mathbb C)$, $\Omega_{+}$ is an Hermitian
$2\times2$ matrix. It is obvious that this matrix can be reduced to the canonic form by
conjugations of $Sl(2,\mathbb C)$. Now, we consider dimension $d=5+1$. The group $Spin(5,1)$
is isomorphic to $SU^{*}(4)$. Hence, $\Omega_{+}$ is an Hermitian $4\times4$ matrix. On the
other hand, it was shown in Sec. 4 that this matrix can be reduced to the canonic form by
conjugations of $Sp(2)$. Since $Sp(2)\subset SU^{*}(4)$, it follows that this reduction is
possible in the considered case. Thus, there exists unique to within equivalence nontrivial
system of BPS equations for any pair of values $d\leq 6$ and $\nu=\nu(d)$.
\par
Now, we will construct these systems. It is obvious that in dimension $d=1+1$, we has only the
vacuum $F_{ab}=0$. We consider dimension $d=3+1$. If follows from (\ref{12-04}) and Table 1
that
\begin{equation}
Cl^0_{3,1}(\mathbb R)\simeq Cl_{3,0}(\mathbb R)\simeq\mathbb C(2).
\end{equation}
Therefore, the subalgebra $Cl_{3,1}^0(\mathbb R)$ decomposes into the direct sum of two
minimal left ideals. We construct the total orthogonal system of primitive idempotents
\begin{equation}
\Omega_{\alpha}=\frac12(1\pm\Gamma_{14}).
\end{equation}
It follows easily that the corresponding system of BPS equations has only the trivial solution
$F_{ab}=0$. We consider dimension $d=5+1$. Since
\begin{equation}
Cl^0_{5,1}(\mathbb R)\simeq Cl_{5,0}(\mathbb R)\simeq\mathbb H(2)\oplus\mathbb H(2),
\end{equation}
it follows that the subalgebra $Cl_{5,1}^0(\mathbb R)$ decomposes into the direct sum of four
minimal left ideals. We construct the total orthogonal system of primitive idempotents
\begin{equation}
\Omega_{\alpha}=\frac14(1\pm\Gamma_{123456})(1\pm\Gamma_{1234}).
\end{equation}
Substituting $\Omega_1$ in the Eq. (\ref{13-02}), we get the BPS equations
\begin{equation}\label{17-02}
\begin{aligned}
F_{ab}&=\frac12\varepsilon_{abcd}F_{cd},\\
F_{a5}&=-F_{a6}.
\end{aligned}
\end{equation}
Conversely, the system of BPS equations constructed with the help of the idempotent
\begin{equation}
\Omega_{12}=\frac12(1+\Gamma_{123456}),
\end{equation}
has only trivial solution. Hence, any nontrivial system of BPS equations in dimension $d=5+1$
defined by the chiral representation of $Spin(5,1)$ is equivalent to the system (\ref{17-02}).
The theorem is proved.

\subsection{BPS states in the supersymmetric Yang-Mills theories}

Now we apply the obtained above results to the supersymmetric Yang-Mills theories. First, we
note that for each choice of the infinitesimal supersymmetry parameter $\varepsilon$, there is
a corresponding conserved supercharge $Q$. Out of this infinity of conserved supercharges, we
wish to identify those that generate unbroken supersymmetries. An unbroken supersymmetry $Q$
is simply a conserved supercharge that annihilates the vacuum state $|\Phi\rangle$. Saying
that $Q$ annihilates $|\Phi\rangle$ is equivalent to saying that for all operators $U$,
$\langle\Phi|\{Q,U\}|\Phi\rangle=0$. This will certainly be so $U$ is a bosonic operator,
since then $\{Q,U\}$ is fermionic, so the real issue is whether
$\langle\Phi|\{Q,U\}|\Phi\rangle$ vanishes when $U$ is a fermionic operator. Now, when $U$ is
fermionic $\{Q, U\}$ is simply $\delta U$, the variation of $U$ under the supersymmetry
transformation generated by $Q$. Also, in the classical limit, $\delta U$ and
$\langle\Phi|\delta U|\Phi\rangle$ coincide. So finding an unbroken supersymmetry at tree
level means finding a supersymmetry transformation such that $\delta U=0$ for every fermionic
field $U$. Also, in the classical limit, it is enough to check this for elementary fermion
fields.
\par
Further, the open superstring theory can be approximated at low energy by a supersymmetric
Yang-Mills theory. Such theories are described by an action of the form
\begin{equation}\label{17-05}
S=\int d^{D}x\left(-\frac14F^2+\frac{i}{2}\bar\psi\Gamma\cdot D\psi\right).
\end{equation}
The supersymmetry transformations that leave (\ref{17-05}) invariant are
\begin{align}
\delta A_{\mu}&=\frac{i}{2}\bar\varepsilon\,\Gamma_{\mu}\psi,\\
\delta\psi&=-\frac14F_{\mu\nu}\Gamma^{\mu\nu}\varepsilon,\label{17-06}
\end{align}
where $\varepsilon$ is a constant anticommuting spinor. It is well known that the
supersymmetric Yang-Mills theories exists only in the $D=3$, 4, 6 and 10. Using Theorem 2, we
prove that the condition $\delta\psi=0$ in $D\leq 6$ is true only if either $F_{\mu\nu}$ is a
solution of (\ref{17-02}) or $F_{\mu\nu}=0$.
\par
We consider the dimension $D=10$. The Majorana-Weyl spinor $\psi$ in $D=10$ has 16 real
components. On shell these components must still satisfy the Dirac equation that relates eight
of them to the other eight. Therefore, if the values of $F_{\mu\nu}$ are arbitrary, then it
follows from (\ref{17-06}) that only eight components of $\varepsilon$ are independent. We
choose an orthonormal basis in $\text{Ker}(F_{\mu\nu}\Gamma^{\mu\nu})\subset V$ and extend it
to the spinor space $V$ so that only eight components of $\varepsilon$ are not zero. Then the
condition $\delta\psi=0$ requires that the projector $\Omega$ in (\ref{13-02}) has the block
diagonal form (\ref{13-12}). It is obvious that it can be reduced to the canonic form by
transformations from $SO(8)\subset SO(9,1)$. Hence, for every value of $\nu$, there exists
unique to within equivalence nontrivial system of BPS equations. In order that to find these
systems we construct the total orthogonal system of primitive idempotents
\begin{equation}
\Omega_{\alpha}=\frac{1}{32}(1\pm\Gamma_{*})(1\pm\Gamma_{12345678})
(1\pm\Gamma_{1234})(1\pm\Gamma_{1256})(1\pm\Gamma_{1476})
\end{equation}
of the Clifford algebra $Cl^0_{9,1}(\mathbb R)$. Substituting $\Omega_1$ in the equation
(\ref{13-02}), we get the following BPS equations
\begin{equation}\label{17-08}
\begin{aligned}
F_{ab}&=\frac12f_{abcd}F_{cd},\\
F_{a9}&+F_{a10}=F_{9\,\!10}=0,
\end{aligned}
\end{equation}
where $f_{abcd}$ is a completely antisymmetric tensor with the components (\ref{17-04}).
Obviously, $\nu=2/32$. The systems of BPS equations for other values of $\nu$ can be obtained
by the method of Sec. 5. Thus, nontrivial state of unbroken supersymmetry in the
supersymmetric Yang-Mills theories there exist only for $D=6$ and 10. Also, in dimension
$D=6$, such state is an instanton solutions of (\ref{17-02}). In dimension $d=10$, such states
are either solutions of the system (\ref{17-08}) or solutions of the BPS equations in eight
dimensions adding the conditions $F_{a9}=F_{a10}=F_{9\,\!10}=0$.

\subsection{Heterotic string solitons}

In conclusion, we discuss the possibility of using the results obtained above to construct
soliton solutions of the low-energy effective theory of the heterotic string. For the
heterotic string, the low-energy effective action is identical to the $D=10$ $N=1$
supergravity and super Yang-Mills action. The bosonic part of this action reads
\begin{equation}
S=\frac{1}{2k^2}\int d^{10}x\,\sqrt{-g}e^{-2\phi}\left(R+4(\nabla\phi)^2
-\frac{1}{3}H^2-\frac{\alpha'}{30}\text{Tr}F^2\right).
\end{equation}
We are interested in solutions that preserve at least one supersymmetry. This requires that in
10 dimensions there exist at least one Majorana-Weyl spinor $\varepsilon$ such that the
supersymmetry variations of the fermionic fields vanish for such solutions
\begin{align}
\delta\chi&=F_{MN}\Gamma^{MN}\varepsilon,\label{17-09}\\
\delta\lambda&=(\Gamma^{M}\partial_{M}\phi-\frac16H_{MNP}\Gamma^{MNP})\varepsilon,
\label{17-10}\\
\delta\psi_{M}&=(\partial_{M}+\frac14\Omega_{M}^{AB}\Gamma_{AB})\varepsilon.\label{17-11}
\end{align}
Here $\phi$ is the dilaton field, $F_{MN}$ is the Yang-Mil1s field strength, and $H$ is the
gauge-invariant field strength of the antisymmetric tensor field $B_{MN}$. While we can
arbitrarily specify the space-time metric and the dilaton field $\phi$ in trying to obey
\begin{equation}
\delta\chi=\delta\lambda=\delta\psi_{M}=0,\label{17-12}
\end{equation}
we cannot arbitrarily specify $F$ or $H$; they must obey certain Bianchi identities. In the
string theory these identities have the form
\begin{equation}\label{17-13}
dH=\alpha'\left(\text{tr} R\wedge R-\frac{1}{30}\text{Tr} F\wedge F\right).
\end{equation}
Note that the connection $\Omega_{M}$ in (\ref{17-11}) is a non-Riemannian. It is related to
the usual spin connection $\omega$ by
\begin{equation}
\Omega_{M}^{AB}=\omega_{M}^{AB}-H_{M}^{AB}.
\end{equation}
\par
The analysis of (\ref{17-09}), (\ref{17-10}), and (\ref{17-11}) is rather complicated in
general, and so we simplify the discussion by assuming at the outset that the Majorana-Weyl
spinor $\varepsilon$ is constant. Further, we suppose that a subgroup $G$ of $SO(9,1)$ is a
group of symmetries of BPS equations, and we choose $\varepsilon$ to be a $G$ singlet of the
Majorana-Weyl spinor. Then, for suitable $G$, there exists a completely antisymmetric tensor
$f_{abcd}$ such that the ansatz
\begin{equation}\label{17-14}
\begin{aligned}
g_{ab}&=e^{\phi}\delta_{ab},\\
H_{abc}&=\lambda f_{abcd}\partial^{d}\phi,
\end{aligned}
\end{equation}
solves the supersymmetry equations with zero background fermi fields provided the Yang-Mills
gauge fields satisfies the BPS equations. Such solutions were found in the works~[14--23]. The
obtained above classification of BPS equations in the Euclidean and Minkowski spaces permits
to describe all such solutions at least with ansatz (\ref{17-14}). It is interestingly that at
present, states of unbroken supersymmetry are very nearly the only examples known of
compactified solutions of the equations; the other known examples are related in comparatively
simple ways to states of unbroken supersymmetry.

\bigskip\medskip\par\noindent
{\bf Acknowledgements}
\medskip\par\noindent
The research was supported by RFBR Grant 06-02-16140.

\end{document}